\documentclass[aps]{revtex4}

\usepackage{amsmath,latexsym,amssymb,amsfonts,mathbbol,amsthm,citesort}

\renewcommand{\v}[1]{\mathbf{#1}}
\newcommand{\C}[1]{\mathcal{#1}}

\renewcommand{\d}[0]{\partial}

\newcommand{\ad}[0]{^{\dag}}

\newcommand{\smallf}[2]{ {\textstyle \frac{#1}{#2}} }

\newcommand{\wt}[1]{\widetilde{#1}}

\newcommand{\EE}[0]{\mathbb{E}}
\newcommand{\RR}[0]{\mathbb{R}}

\newcommand{\ZZ}[0]{\mathbb{Z}}

\newcommand{\NN}[0]{\mathbb{N}}

\renewcommand{\Re}[0]{\mathfrak{Re}\!~}

\newcommand{\lat}[0]{\mathsf{lat}}

\newlength{\minus}
\newcommand{\ms}[0]{\hspace{\minus}}

\newenvironment{mat}[1]{\begin{array}{@{}*{#1}{r@{}l}@{}}}{\end{array}}

\DeclareMathOperator{\diag}{diag}

\theoremstyle{plain}
\newtheorem{theorem}[subsection]{Theorem}

\begin{document}

\title{Compactification along lightlike lattices}
\author{Hanno Hammer}
\email{H.Hammer@umist.ac.uk}
\affiliation{UMIST, Department of Mathematics \\ PO Box 88 \\
Manchester M60 1QD \\ United Kingdom}
\date{\today}

\begin{abstract}
Spacetimes obtained by dimensional reduction along lattices containing
a lightlike direction can admit semigroup extensions of their isometry
groups. We show by concrete examples that such a semigroup can exhibit
a natural order, which in turn implies the existence of preferred
coordinate charts on the underlying space. Specifically, for
spacetimes which are products of an external Minkowski space with an
internal two-dimensional Lorentzian space, where one of the lightlike
directions has a compact size, the preferred charts consist of
"infinite-momentum" frames on the internal space. This implies that
fields viewed from this preferred frame acquire extreme values; in
particular, some of the off-diagonal components of the
higher-dimensional metric, which may be regarded as gauge potentials
for a field theory on the external Minkowski factor, vanish. This
raises the possibility of regarding known gauge theories as part of
more extended field multiplets which have been reduced in size since
they are perceived from within an extreme frame. In the case of an
external 4-dimensional Minkowski spacetime times a two-dimensional
Lorentzian cylinder, the field content as seen in the preferred frame
is that of a five-dimensional Kaluza-Klein theory, where the
electrodynamic potentials $A_{\mu}$ may depend, in addition to the
external spacetime coordinates, on a fifth coordinate along a
lightlike direction. The fact that the metric along this direction is
zero obstructs the generation of field equations from the Ricci tensor
of the overall metric.
\end{abstract}

\maketitle


\section{Introduction}

Spacetimes which are obtained by dimensional reduction along lightlike
directions can exhibit peculiarities in their geometrical properties
which are not found in other compactifications. The original
motivation for the work presented in this paper has been to examine
the structure of the isometry group of flat compactified spacetimes
when dimensional reduction is performed over a lattice containing a
lightlike direction. In the course of our investigation we have
uncovered some interesting results about the geometry of
higher-dimensional spacetimes $M$ which are product manifolds with an
"internal factor" which contains a lightlike circle. It turns out that
such spaces can admit semigroup extensions $e I(M)$ of their isometry
group $I(M)$; that is to say, they admit smooth global maps $\Lambda:
M \rightarrow M$ which are surjective, but no longer injective on $M$,
{\it and which locally preserve the metric}, so that they still
qualify as "isometries", albeit in a wider sense. Any two such
transformations $g$ and $g'$ again combine to a, generally
non-invertible, metric-preserving transformation $g g'$; but the
non-injectiveness implies that global inverses of such maps do not
generally exist. This fact accounts for the semigroup structure of
these maps.

Such a semigroup extension suggests interesting consequences for the
existence of preferred coordinate charts on the space underlying; for,
a semigroup of isometries can be given a natural ordering on account
of the fact that the product $g g' = h$ of two semigroup elements $g$
and $g'$ still belongs to $eI(M)$, but there may be no way to resolve
this product for either $g$ or $g'$. Below we shall walk through the
simplest example possible, namely a cylindrical internal spacetime
where the circle of the cylinder is lightlike; we shall see that in
this case the preferred coordinate charts consist of
"infinite-momentum" frames, i.e., coordinate frames which are related
to any other frame by the limit of a series of internal Lorentz
transformations, i.e., an infinite boost. Viewed from such a frame,
fields defined on the spacetime acquire extreme values; in particular,
some of the off-diagonal components of the higher-dimensional metric,
which may be regarded as gauge potentials for a field theory on the
$(3+1)$-dimensional external spacetime factor, vanish. This raises the
possibility of regarding known gauge theories, with a given number of
gauge potentials, as part of a more extended field multiplet, which
has been "reduced" in size since it is perceived from within an
"extreme" frame. What is more, the semigroup structure implies that
all fields defined on the internal spacetime factor must be
independent of one of the lightlike coordinates on this factor, namely
the one which has acquired a compact size in the process of
dimensional reduction. We then end up being able to explain quite
naturally how one dimension on the internal manifold must cease to
reveal its presence, since all fields must be independent of this
dimension. Furthermore, the geometry of the lightlike cylinders
studied below has another interesting property: This property is given
by the fact that all lightlike cylinders, independent of the size of
the "compactification radius" along the lightlike direction, are {\it
isometric}. This implies that any physical theory which is formulated
covariantly in terms of geometric objects on the higher-dimensional
spacetime is automatically independent of the compactification radius
along one of the lightlike directions. This then means that the
question whether this radius is microscopic or large becomes
irrelevant, since no covariant theory can distinguish between two
different radii!

Below we shall work out explicitly the example of a $6$-dimensional
spacetime which is a product of a $4$-dimensional Minkowski spacetime
and an internal lightlike cylinder. In this case the field content as
seen in the "infinite-momentum frame" on the internal space is that of
a $5$-dimensional Kaluza-Klein theory, where the electrodynamic
potentials $A_{\mu}$ may depend, in addition to the four external
spacetime coordinates, on a fifth coordinate along a lightlike
direction. We discuss some of the problems associated with obtaining
meaningful equations of motion for these potentials, which formally
are equations on a manifold with {\it degenerate metric}. This leads
outside the established framework of semi-Riemannian geometry, and
hence it is not clear how to produce dynamical equations for the
potentials $A_{\mu}$.

Lightlike compactification, or "discretization", has been utilized
within the framework of Discrete Light-Cone Quantization, a scheme
which has been proposed for gauge field quantization \cite{Thorn1978}
and a lattice approach to string theory \cite{GilesThorn1977}; it goes
back to Dirac's original idea of light-cone quantization
\cite{DiracLightCone} which was later rediscovered by Weinberg
\cite{Weinberg1966a}. While in Discrete Light-Cone Quantization
lightlike directions are compactified (usually one or two), it is
assumed that these directions are obtained from a coordinate
transformation onto an infinite-momentum reference frame living in
ordinary 3+1-dimensional Minkowski spacetime. In contrast, our work
focuses on the compactification of lightlike directions in a
higher-dimensional covering spacetime. Geometrical aspects of
dimensional reduction along lightlike Killing vectors, and action
principles in such spaces, were investigated in
\cite{JuliaNicolai1995a}. These authors studied compactifications
along continuous orbits; in contrast, in this work we are dealing with
orbits which result from the action of discrete groups.

The plan of the paper is as follows: In section \ref{OrbitSpaces} we
motivate the emergence of semigroup extensions from a more abstract
point of view by considering how isometry groups of orbit spaces are
obtained from isometry groups of associated covering spaces. In
section \ref{IdOverLattice} we discuss a condition that obstructs the
existence of semigroup extensions for orbit spaces which are obtained
from the action of primitive lattice translations of a suitable
lattice on a flat pseudo-Euclidean covering space. In section
\ref{2DGeometry} we study in detail the example of two-dimensional
Lorentzian cylindrical spaces which are obtained by compactifying a
lightlike direction in a Lorentzian covering space. We show that the
associated semigroup extension of the isometry group of the cylinder
exhibits a natural ordering, which may be interpreted as pointing out
preferred, "infinite-momentum", frames on this space. The consequences
of these semigroup transformations for scalars, vectors and covectors
on the cylinder are examined in sections \ref{Scalar} and
\ref{Vectors}. In section \ref{KaluzaKlein} we regard the cylindrical
space as the "internal" factor in a product spacetime whose external
factor is Minkowski. We show that the preferred charts on the total
space reduce the associated, originally $(4+2)$-dimensional,
Kaluza-Klein theory to an effective $(3+1+0)$-dimensional Kaluza-Klein
scenario, where in the latter case the internal space appears
one-dimensional and carries a zero metric. In \ref{Summary} we
summarize our results.

\section{Orbit spaces and normalizing sets \label{OrbitSpaces}}

We have mentioned in the introduction that semigroup extensions of
isometry groups emerge naturally when dimensional reduction along
lightlike directions is performed. In order to see the key point at
which such extensions present themselves it is best to approach the
subject in a more general way, by motivating the idea of the {\it
extended normalizer} of a set of symmetry transformations. To this end
we start by recalling how symmetry groups, and possible extensions
thereof, of identification spaces are obtained from the symmetry
groups of associated covering spaces. The relevant idea here is that
of diffeomorphisms on a covering space which {\it descend to a
quotient space}:

Let $M$ be a connected pseudo-Riemannian manifold with a metric $\eta
$, and let $I(M)$ be the group of isometries of $M$. Assume that a
discrete subgroup $\Gamma \subset I(M)$ acts properly discontinuously
and freely on $M$; in this case the natural projection $p : M
\rightarrow M / \Gamma$ from $M$ onto the space of orbits, $M / \Gamma
$, can be made into a covering map, and $M$ becomes a covering space
of $M / \Gamma$ (e.g.  \cite{Brown,Fulton,Jaehnich,Massey}). In fact
there is a unique way to make the quotient $M / \Gamma$ a
pseudo-Riemannian manifold (e.g.  \cite
{Wolf,Poor,SagleWalde,Warner,ONeill}): In this construction one
stipulates that the projection $p$ be a local isometry, which
determines the metric on $M / \Gamma$. In such a case, we call $p : M
\rightarrow M / \Gamma $ a pseudo-Riemannian covering.

In any case the quotient $p: M \rightarrow M / \Gamma$ can be regarded
as a fibre bundle with bundle space $M$, base $M/ \Gamma$, and
$\Gamma$ as structure group, the fibre over $m \in M/\Gamma$ being the
orbit of any element $x \in p^{-1}(m)$ under $\Gamma $,
i.e. $p^{-1}(m) = \Gamma x = \left\{ \gamma x \mid \gamma \in \Gamma
\right\}$. If $g \in I(M)$ is an isometry of $M$, then $g$ gives rise
to a well-defined map $g_{\#}: M / \Gamma \rightarrow M / \Gamma $,
defined by
\begin{equation}
\label{map1}
 g_{\#}(\Gamma x) \equiv \Gamma(gx) \quad,
\end{equation}
on the quotient space {\bf only} when $g$ preserves all fibres, i.e.
when $g \left(\Gamma x \right) \subset \Gamma(gx)$ for all $x\in M$.
This is equivalent to saying that $g \Gamma g^{-1} \subset \Gamma$. If
this relation is replaced by the stronger condition $g \Gamma g^{-1} =
\Gamma$, then $g$ is an element of the {\it normalizer}
$N\left(\Gamma\right)$ of $\Gamma$ in $I(M)$, where
\begin{equation}
\label{normalizer1}
 N \left( \Gamma \right) = \left\{ g \in I(M) \mid g \Gamma g^{-1} =
  \Gamma \right\} \quad .
\end{equation}
The normalizer is a group by construction. It contains all
fibre-preserving elements $g$ of $I(M)$ such that $g^{-1}$ is
fibre-preserving as well. In particular, it contains the group
$\Gamma$, which acts trivially on the quotient space; this means that
for any $\gamma \in \Gamma$, the induced map $\gamma_{\#} : M / \Gamma
\rightarrow M / \Gamma$ is the identity on $M / \Gamma $. This
follows, since the action of $\gamma_{\#}$ on the orbit $\Gamma x$,
say, is defined to be $\gamma_{\#} \left( \Gamma x \right) \equiv
\Gamma \left( \gamma x \right) = \Gamma x$, where the last equality
holds, since $\Gamma$ is a group.

In this work we are interested in relaxing the equality in the
condition defining $N\left(\Gamma\right)$; to this end we introduce
what we wish to call the {\it extended normalizer}, denoted by
$eN\left(\Gamma\right)$, through
\begin{equation}
\label{exNorm}
 eN(\Gamma) \equiv \left\{ g \in I(M) \mid g \Gamma g^{-1} \subset
 \Gamma \right\} \quad .
\end{equation}
The elements $g \in I(M)$ which give rise to well-defined maps
$g_{\#}$ on $ M / \Gamma$ are therefore precisely the elements of the
extended normalizer $eN(\Gamma)$, as we have seen in the discussion
above. Such elements $g$ are said to {\it descend to the quotient
space} $M / \Gamma$.  Hence $eN(\Gamma)$ contains all isometries of
$M$ that descend to the quotient space $M / \Gamma$; the normalizer
$N(\Gamma )$, on the other hand, contains all those $g$ for which
$g^{-1}$ descends to the quotient as well. Thus, $N(\Gamma)$ is the
group of all $g$ which descend to {\bf invertible} maps $g_{\#}$ on
the quotient space. In fact, the normalizer $N(\Gamma)$ contains all
isometries of the quotient space, the only point being that the action
of $N(\Gamma)$ is not effective, since $\Gamma \subset N( \Gamma)$
acts trivially on $M / \Gamma$.  However, $\Gamma$ is a normal
subgroup of $N (\Gamma)$ by construction, so that the quotient
$N(\Gamma) / \Gamma$ is a group again, which is now seen to act
effectively on $M / \Gamma$, and the isometries of $M / \Gamma$ which
descend from isometries of $M$ are in a 1--1 relation to elements of
this group. Thus, denoting the isometry group of the quotient space
$M/\Gamma$ as $I\left(M/\Gamma\right)$, we have the well-known result
that
\begin{equation}
\label{normaliz1}
 I (M/ \Gamma) = N (\Gamma) / \Gamma \quad .
\end{equation}

Now we turn to the extended normalizer. For an element $g \in
eN(\Gamma)$, but $g \not \in N(\Gamma)$, the induced map $g_{\#}$ is
no longer injective on $M / \Gamma$: To see this we observe that now
the inclusion in definition (\ref{exNorm}) is proper, i.e.  $g \Gamma
g^{-1} \subsetneqq \Gamma $. It follows that a $\gamma' \in \Gamma$
exists for which
\begin{equation}
\label{hilf2}
g \gamma g^{-1} \neq \gamma' \quad \text{for all $\gamma \in \Gamma$}
\quad.
\end{equation}
Take an arbitrary $x \in M$; we claim that
\begin{equation}
\label{hilf3}
 g \Gamma g^{-1} x \subsetneqq \Gamma x \quad.
\end{equation}
To see this, assume to the contrary that the sets $g \Gamma g^{-1} x$
and $\Gamma x$ coincide; then a $\gamma_1 \in \Gamma$ exists for which
$g \gamma_1 g^{-1} x = \gamma' x$; since $g$ is an element of the
extended normalizer $eN(\Gamma)$, $g \gamma_1 g^{-1} \in \Gamma$,
i.e., $g \gamma_1 g^{-1} = \gamma_2$, say. It follows that $\gamma_2 x
= \gamma' x$ or $\gamma_2^{-1} \gamma' x = x$. The element
$\gamma_2^{-1} \gamma'$ belongs to $\Gamma$, which, by assumption,
acts freely. Free action implies that if a group element has a fixed
point then it must be the unit element, implying that $\gamma' =
\gamma_2 = g \gamma_1 g^{-1}$, which contradicts (\ref{hilf2});
therefore, (\ref{hilf3}) must hold. Eq. (\ref{hilf3}) can be expressed
by saying that the orbit of $x$ is the image of the orbit of $g^{-1}
x$ under the action of the induced map $g_{\#}$, $g_{\#} (\Gamma
g^{-1} x) = \Gamma x$; but that the $g_{\#}$-image of the orbit
$\Gamma g^{-1} x$, regarded as a set, is properly contained in the
orbit of $x$. The latter statement means that a $\gamma' \in \Gamma$
exists, as in (\ref{hilf2}), such that $\gamma' x \neq g \gamma g^{-1}
x$ for all $\gamma$. It follows that $\gamma g^{-1} x \neq g^{-1}
\gamma' x$ for all $\gamma$, implying that the point $g^{-1} \gamma'
x$ is not contained in the orbit of the point $g^{-1} x$. Its own
orbit, $\Gamma g^{-1} \gamma' x$, is therefore distinct from the orbit
$\Gamma g^{-1} x$ of the point $g^{-1} x$, since two orbits either
coincide or are disjoint otherwise. However, the induced map $g_{\#}$
maps $\Gamma g^{-1} \gamma' x$ into
\begin{equation}
\label{hilf4}
 g_{\#} (\Gamma g^{-1} \gamma' x) = \Gamma( g g^{-1} \gamma' x) =
 \Gamma x \quad,
\end{equation}
from which it follows that $g_{\#}$ maps two distinct orbits into the
same orbit $\Gamma x$, expressing the fact that $g_{\#}$ is not
injective. In particular, if $g$ was an isometry of $M$, then $g_{\#}$
can no longer be a global isometry on the quotient space, since it is
not invertible {\bf on the quotient space}. From this fact, or
directly from its definition (\ref{exNorm}), we infer that the
extended normalizer is a semigroup, since it contains the identity,
and with any two elements $g$ and $g'$ also their product $g g'$. On
the other hand, we shall see below that there are cases where the
elements $g$ of the extended normalizer are still locally injective,
in particular, injective on the tangent spaces of the
compactification; and moreover, they preserve the metric on the
tangent spaces, so that they should be regarded as a kind of
"generalized isometries".

\section{Identifications over lattices \label{IdOverLattice}}

So far we have been entirely general with regard to the manifolds
$M$. We now make more specific assumptions: Our covering spaces $M$
are taken to be a flat pseudo-Euclidean space $\RR_t^n$, i.e., $\RR^n$
endowed with a pseudo-Euclidean metric $\eta$ with signature $(-t,+s),
t+s=n$. The isometry group $I(\RR^n_t)$ of this space is a semidirect
product $\RR^n \odot O(t,n-1)$, where $\RR^n$ denotes the
translational factor, and will be denoted by $\EE(t,n-t)$. The
identification group $\Gamma$ will be taken as the set of primitive
lattice transformations of a lattice in $\RR^n_t$. If the $\RR$-linear
span of the lattice vectors has real dimension $m$, say, the resulting
identification space $M / \Gamma$ is homeomorphic to a product
manifold $\RR^{n-m} \times T^m$, where $T^m$ denotes an
$m$-dimensional torus. This space inherits the metric from the
covering manifold $\RR^n_t$, since the metric is a local object, but
the identification changes only the global topology. Thus $M / \Gamma$
is again a manifold, but may cease to be semi-Riemannian, since the
metric on the torus may turn out to be degenerate. Whereas the
isometry group of the covering space $\RR^n_t$ is $\EE(t,n-t)$, the
isometry group of the "compactified" space $M / \Gamma$ is obtained
from the normalizer $N(\Gamma)$ of the group $\Gamma$ in $\EE(t,n-t)$
according to formula (\ref{normaliz1}).

For our purposes it is sufficient to consider lattices that contain
the origin $0 \in \RR_t^n$ as a lattice point. Let $1~\le~m~\le~n$,
let $\underline{u} \equiv (u_1,\ldots ,u_m)$ be a set of $m$ linearly
independent vectors in $\RR_t^n$; then the $\ZZ$-linear span of
$\underline{u}$,
\begin{equation}
\label{latpoints1}
 \mathsf{lat} \equiv \Big\{ \sum_{i=1}^m z_i \cdot u_i \Big| z_i \in
\mathbb{Z} \Big\} \quad ,
\end{equation}
is the set of {\it lattice points} with respect to
$\underline{u}$. Let $U$ denote the $\RR$-linear span of
$\underline{u}$, or equivalently, of $\mathsf{lat}$.

The subset $T_{\mathsf{lat}}\subset \EE(t,n-t)$ is the subgroup of all
translations in $\EE(t,n-t)$ through elements of $\mathsf{lat}$,
\begin{equation}
\label{pp3fo5}
 T_{\mathsf{lat}}=\big\{(t_z,1)\in \EE(t,n-t) \big| t_z\in
 \mathsf{lat} \big\} \quad .
\end{equation}
Elements $(t_z,1)$ of $T_{\mathsf{lat}}$ are called {\it primitive
lattice translations}. $T_{\mathsf{lat}}$ is taken as the discrete
group $\Gamma$ of identification maps, which gives rise to the
identification space $p: \RR^n_t \rightarrow \RR^n_t / T_{\lat}$.

We now examine the normalizer and extended normalizer of the
identification group: For an element $(t,R) \in \EE(t,n-t)$ to be in
the extended normalizer $eN(T_{\lat})$ of $T_{\lat}$, the condition
\begin{equation}
 (t,R)(t_z,1)(t,R)^{-1} = (Rt_z,1) \in T_{\mathsf{lat}}
\label{Beding1}
\end{equation}
must be satisfied for lattice vectors $t_z$. In other words, $Rt_z \in
\lat$, which means that the pseudo-orthogonal transformation
$R$ must preserve the lattice $\lat$, $R \lat \subset
\lat$. For an element $(t,R)$ to be in the normalizer,
$(t,R)^{-1}$ must be in the normalizer as well, implying $R^{-1}
\lat \subset \lat$, so altogether $R \lat =
\lat$. The elements $R$ occuring in the normalizer therefore
naturally form a subgroup $G_{\lat}$ of $O(t,n-t)$; on the
other hand, the elements $R$ occuring in the extended normalizer form
a {\it semigroup} $eG_{\lat}\supset G_{\lat}$.
Furthermore, no condition on the translations $t$ in $(t,R)$ arises,
hence all translations occur in the normalizer as well as in the
extended normalizer. Thus, the [extended] normalizer has the structure
of a semi-direct [semi-]group
\begin{subequations}
\label{Norm1}
\begin{align}
 N\left( \Gamma \right) & = \mathbb{R}^n\odot G_{\lat} \quad ,
 \label{Norm1a} \\
 eN\left( \Gamma \right) & = \mathbb{R}^n\odot eG_{\lat}
 \quad, \label{Norm1b}
\end{align}
\end{subequations}
where $\RR^n$ refers to the subgroup of all translations in
$\EE(t,n-t)$.

A sufficient condition under which $eG_{\lat}$ coincides with
$G_{\lat}$ is easily found:

\begin{theorem}[Condition] \label{Bedingung}
  If the restriction $\eta|U$ of the metric $\eta$ to the real linear
  span $U$ of the lattice vectors is positive- or negative-definite,
  then $eG_{\mathsf{lat}} = G_{\mathsf{lat}}$.
\end{theorem}

{\it Proof:}

We start with the case that $\eta|U$ is positive definite. Assume that
$R \in eG_{\lat} \subset O(t,n-t)$, but $R \not\in G_{\lat}$. Then $R$
preserves the lattice but is not surjective, i.e., $R \lat \subsetneqq
\lat$. This implies the series of inclusions
\begin{equation}
\label{incl1}
 \cdots R^3 \lat \subsetneqq R^2 \lat \subsetneqq R\lat \subsetneqq
 \lat \quad.
\end{equation}
Now choose an $x \in \lat \backslash R \lat$; then it follows that $R
x \in R \lat \backslash R^2 \lat, \ldots, R^k x \in R^k \lat
\backslash R^{k+1} \lat$. The series of proper inclusions
(\ref{incl1}) then implies that the elements in the series $k \mapsto
R^k x$ are all distinct from each other; furthermore, they are all
lattice points, since $R$ preserves the lattice. We then see that the
set of lattice points $\big\{\, R^k x\, \big|\, k \in \NN_0\, \big\}$
must be infinite. This means that there are elements in this set whose
Euclidean norm $\sqrt{ (\eta|U)(R^k x, R^k x)}$ exceeds every finite
bound. However, the transformations $R$ are taken from the overall
metric-preserving group $O(t,n-t)$ and hence must also preserve the
restricted metric $\eta|U$. The latter statement implies that
\begin{equation}
\label{incl2}
 \cdots (\eta|U)(R^3 x, R^3 x) = (\eta|U)(R^2 x, R^2 x) = \eta(Rx,Rx) =
 \eta(x,x) \quad,
\end{equation}
hence all elements $R^k x$ have the same norm according to
(\ref{incl2}). This contradicts the previous conclusion that there are
elements for which the norm exceeds all bounds. This shows that our
initial assumption $R \not \in G_{\lat}$ was wrong.

If $\eta|U$ is assumed to be negative-definite, the argument given
above clearly still applies, since the norm of an element $x$ in this
case is just given by $\sqrt{-(\eta|U)(x,x)}$. This completes our
proof.  {\hfill$\blacksquare$}

We see that the structure of the proof relies on the fact that the
restricted metric $\eta|U$ was Euclidean. If this metric were
pseudo-Euclidean instead, the possibility that $eG_{\lat} \supsetneqq
G_{\lat}$ arises. We shall now study an explicit example of this
situation.

\section{Two-dimensional cylinders with a compact lightlike direction
\label{2DGeometry} }

We now study a specific example of a space, or rather, a family of
spaces, for which natural semigroup extensions can be derived from the
isometry group of a simply-connected covering space.

The quotient spaces under consideration are two-dimensional cylinders
and will be denoted by $C^2_r$; they are defined as follows:
\begin{equation}
\label{def1}
 C^2_r \equiv [0,r) \times \RR \quad, \quad r \sim 0 \quad,
\end{equation}
i.e., $[0,r)$ is a circle with perimeter $r >0$, and we have the
identification $0 \sim r$. These spaces shall have canonical
coordinates $(x^+, x^-)$, where $0 \le x^+ < r$ and $x^- \in \RR$. To
the coordinates $(x^+, x^-)$ we shall also refer as {\it lightlike},
for reasons which will become clear shortly. Each of the cylinders
$C^2_r$ is endowed with a metric $h$ which, in lightlike coordinates,
is given as
\begin{equation}
\label{def2}
 h = - dx^+ \otimes dx^- - dx^- \otimes dx^+ \quad.
\end{equation}
The manifolds $C^2_r$ have a remarkable property: They
are all isometric, the isometry $\phi_{rr'} : C^2_r \rightarrow
C^2_{r'}$ being given by
\begin{equation}
\label{iso1}
 \phi_{rr'}\big(\, x^+, \, x^-\, \big) = \big(\, x^{\prime +}, \, x^{\prime -}
 \, \big) \equiv \bigg(\, \frac{r'}{r}\, x^+\, ,\; \frac{r}{r'}\, x^-
 \, \bigg) \quad,
\end{equation}
where $(x^+, x^-)$, $(x^{\prime +}, x^{\prime -})$ are lightlike
coordinates on $C^2_r$ and $C^2_{r'}$, respectively. Since the metric
on $C^2_{r'}$ is given by
\begin{equation}
\label{iso2}
 h' = - dx^{\prime +} \otimes dx^{\prime -} - dx^{\prime -} \otimes
 dx^{\prime +} \quad,
\end{equation}
we see immediately that $h$ and $h'$ are related by pull-back,
\begin{equation}
\label{iso3}
 \phi^*_{rr'}\, h' = h \quad.
\end{equation}
This means that the geometry of $C^2_r$, and subsequently, any physics
built upon covariant geometrical objects on $C^2_r$, must be
completely insensitive to the size of the "compactification radius"
$r/2\pi$. This may suggest that it makes sense to consider the limit
$r \rightarrow 0$.

We now show how the cylinders $C^2_r$ are obtained from a
simply-connected two-dimensional covering space: The uncompactified
covering space is taken to be $\RR^2_1$, which is homeomorphic to
$\RR^2$ with canonical coordinates $X \equiv (x^0, x^1)$ with respect
to the canonical basis $(\v{e}_0, \v{e}_1)$, while the metric in this
basis is $\diag(-1,1)$.  The isometry group is $\EE(1,1) = \RR^2 \odot
O(1,1)$, where $\RR^2$ acts as translational subgroup $(t,1)$
according to
\begin{equation}
\label{transsub1}
 \RR^2 \ni t = \left( \begin{mat}{1} & t^0 \\ & t^1 \end{mat} \right)
\quad, \quad (t,1)\, \left( \begin{mat}{1} & x^0 \\ & x^1 \end{mat}
\right) = \left( \begin{mat}{1} & x^0 + t^0 \\ & x^1 + t^1 \end{mat}
\right) \quad.
\end{equation}
A faithful $3$-dimensional real matrix representation of
$\EE(1,1)$ is given by
\begin{equation}
\label{matrep1}
 \left( t, \Lambda \right) = \left( \begin{mat}{2} & \Lambda &&t \\
 &0&& 1
 \end{mat} \right) \quad,
\end{equation}
satisfying the Poincar\'e group law $(t,\Lambda) (t', \Lambda') =
(\Lambda t'+t, \Lambda \Lambda')$. If $\Lambda$ lies in the identity
component of $O(1,1)$ then
\begin{equation}
\label{2d1}
 \Lambda = \left( \begin{mat}{2} & \cosh \alpha && \sinh \alpha \\ &
 \sinh \alpha && \cosh \alpha \end{mat} \right) \quad, \quad \alpha
 \in \RR \quad.
\end{equation}
This representation acts on the coordinates $X\in \RR^2_1$ according
to
\begin{equation}
\label{CoordAct1}
 (t,\Lambda)\, X = \left( \begin{mat}{2} & \Lambda&& t \\ &0&& 1
 \end{mat} \right) \left( \begin{mat}{1} &X \\ &1 \end{mat} \right) =
 \left( \begin{array}{@{}c@{}} \Lambda X + t \\ 1 \end{array} \right)
 \quad.
\end{equation}

We now introduce a one-dimensional lattice with primitive lattice
vector $r \, \v{e}_+$, where $\v{e}_+ \equiv \smallf{1}{\sqrt{2}}
\left( \v{e}_0 + \v{e}_1 \right)$, so that the set $\underline{u}$ as
defined in section \ref{IdOverLattice} contains just one element,
\begin{equation}
\label{lattvec1}
 \underline{u} = \left\{r\, \v{e}_+ \right\} \quad.
\end{equation}
We can introduce lightlike coordinates
\begin{equation}
\label{coord1}
 \left( \begin{mat}{1} & x^+ \\ & x^- \end{mat} \right) =
 \frac{1}{\sqrt{2}} \left( \begin{mat}{2} & 1 && 1 \\ & 1 & - & 1
 \end{mat} \right) \left( \begin{mat}{1} & x^0 \\ & x^1 \end{mat}
 \right) = M\, \left( \begin{mat}{1} & x^0 \\ & x^1 \end{mat} \right)
 \quad,
 \quad M = M^T = M^{-1} \quad,
\end{equation}
on the covering space $\RR^2_1$, so that $\v{e}_+$ is the basis vector
in the $x^+$-direction. The set $\mathsf{lat}$ as defined in
(\ref{latpoints1}) is now given as
\begin{equation}
\label{2d2}
 \mathsf{lat} = \left\{\v{0}, \, \pm r\, \v{e}_+, \, \pm 2r\, \v{e}_+,
 \, \ldots \right\} \quad.
\end{equation}
The subset $T_{\mathsf{lat}}$ as defined in (\ref{pp3fo5}) is now
\begin{equation}
\label{transsub2}
 T_{\mathsf{lat}} = \bigg\{\; (k\, r\, \v{e}_+,\, 1) \in \EE(1,1)\;
\bigg| \; k \in \ZZ \; \bigg\} \quad.
\end{equation}
The elements of $T_{\mathsf{lat}}$ are the primitive lattice
translations, and the identification group is $\Gamma =
T_{\mathsf{lat}}$. After taking the quotient of $\RR^2_1$ over
(\ref{transsub2}), we obtain the two-dimensional cylinder $C^2_r$ as
defined in (\ref{def1}).

According to (\ref{Beding1}), a general Poincare transformation
$(t,\Lambda)$ lies in the extended normalizer of $T_{\mathsf{lat}}$ if
and only if $\Lambda$ preserves the lattice (\ref{2d2}), which is
equivalent to the condition that
\begin{equation}
\label{cond1}
 \Lambda \v{e}_+ = k \cdot \v{e}_+ \quad \text{for some integer $k$ .}
\end{equation}
To facilitate computations we now transform everything into lightlike
coordinates (\ref{coord1}). Then the metric $h= \diag(-1,1)$ takes the
form
\begin{equation}
\label{coord2}
 h = \left( \begin{mat}{2} & 0 & -& 1 \\ -& 1& & 0 \end{mat} \right)
 \quad,
\end{equation}
while a Lorentz transformation (\ref{2d1}) in this basis looks like
\begin{equation}
\label{coord3}
 \Lambda = \left( \begin{mat}{2} & e^{\alpha} && 0 \\ &0& &e^{-\alpha}
 \end{mat} \right) \quad.
\end{equation}
If we allow $\Lambda$ to take values in the full group $O(1,1)$,
rather than the identity component, we obtain matrices
\begin{equation}
\label{coord4}
 \Lambda \quad = \quad \left( \begin{mat}{2} & e^{\alpha} && 0 \\ &0&
 &e^{-\alpha} \end{mat} \right) \quad, \quad - \left( \begin{mat}{2} &
 e^{\alpha} && 0 \\ &0& &e^{-\alpha} \end{mat} \right) \quad, \quad
 \left( \begin{mat}{2} &0 && e^{-\alpha} \\ & e^{\alpha} & & 0
 \end{mat} \right) \quad, \quad - \left( \begin{mat}{2} &0 &&
 e^{-\alpha} \\ & e^{\alpha} & & 0
 \end{mat} \right) \quad.
\end{equation}
Here we have introduced the parity transformation $\C{P}$, defined by
$\C{P}(x^0, x^1) = (x^0, -x^1)$, which acts on lightlike coordinates
according to
\begin{equation}
\label{parity1}
 \C{P} \left( \begin{mat}{1} &x^+ \\ & x^- \end{mat} \right) = \left(
 \begin{mat}{1} & x^{\prime +} \\ & x^{\prime -} \end{mat} \right) = \left(
 \begin{mat}{1} & x^- \\ & x^+ \end{mat} \right) \quad,
\end{equation}
and the time-reversal transformation $\C{T}(x^0, x^1) = (-x^0, x^1) =
- \C{P}(x^0, x^1)$, which acts on lightlike coordinates as
\begin{equation}
\label{timerev1}
  \C{T} \left( \begin{mat}{1} &x^+ \\ & x^- \end{mat} \right) = \left(
 \begin{mat}{1} & x^{\prime +} \\ & x^{\prime -} \end{mat} \right) = \left(
 \begin{mat}{1} -& x^- \\ -& x^+ \end{mat} \right) \quad.
\end{equation}
Parity and time-reversal are the discrete isometries of the metric
(\ref{coord2}). The sequence of matrices in (\ref{coord4}) can then be
written as
\begin{equation}
\label{coord4A}
 \Lambda \quad, \quad \C{P} \C{T} \Lambda = \Lambda \C{P} \C{T} \quad,
 \quad \C{P} \Lambda \quad, \quad \C{T} \Lambda \quad.
\end{equation}
The condition (\ref{cond1}) can be satisfied only for matrices of the
first two kinds; these must have the form
\begin{equation}
\label{coord5}
 \Lambda_m = \left( \begin{mat}{2} &m&&0 \\ &0 & &\frac{1}{m}
 \end{mat} \right) \quad, \quad m \in \ZZ \quad.
\end{equation}
This set of matrices (\ref{coord5}) comprises the group
$eG_{\mathsf{lat}}$ defined in section \ref{IdOverLattice}; it is
obviously a semigroup with composition and unit element
\begin{equation}
\label{coord6}
 \Lambda_m\, \Lambda_{m'} = \Lambda_{m\, m'} \quad, \quad \Eins =
 \Lambda_{1} \quad.
\end{equation}
This semigroup is isomorphic to the set $(\ZZ, \cdot)$ of all integers
with multiplication as composition and one as unit element. For $m
\neq 1$, the inverse of a matrix $\Lambda_m$ in the full Lorentz
group $O(1,1)$ is given by
\begin{equation}
\label{coord7}
 \Lambda_m^{-1} = \Lambda_{1/m} = \left(
 \begin{mat}{2} & \frac{1}{m} &&0 \\ &0&& m \end{mat} \right) \quad.
\end{equation}
However, these inverses violate the condition (\ref{cond1}) and hence
cannot be elements of $eG_{\mathsf{lat}}$; this fact accounts for the
semigroup structure of (\ref{coord5}). Consequently, the only Lorentz
transformation (\ref{coord5}) which preserves the lattice (\ref{2d2})
such that its inverse preserves the lattice as well is the unit matrix
$\Eins$; as a consequence, the group $G_{\mathsf{lat}} = \{ \Eins \}$
is trivial. Then (\ref{Norm1}) takes the form
\begin{subequations}
\label{Norm2}
\begin{align}
 N\left( \Gamma \right) & \simeq \mathbb{R}^2 \quad, \label{Norm2a} \\
 eN\left( \Gamma \right) & \simeq \mathbb{R}^2 \odot (\ZZ, \cdot)
 \quad. \label{Norm2b}
\end{align}
\end{subequations}
In order to obtain the isometry group $I(C^2_r)$, and its semigroup
extension $eI(C^2_r)$, we must divide out the lattice translations
(\ref{transsub2}). Then, in lightlike coordinates (\ref{coord1}),
\begin{subequations}
\label{Norm3}
\begin{align}
 I(C^2_r) & \simeq \bigg\{\; \left( \begin{mat}{1} &t^+ \\ & t^-
 \end{mat} \right) \; \bigg| \; t^+ \in [0,1) \;,\; t^- \in \RR \;
 \bigg\} \quad, \label{Norm3a} \\
 eI(C^2_r) & \simeq \bigg\{\; \left( \begin{mat}{1} &t^+ \\ & t^-
 \end{mat} \right) \; \bigg| \; t^+ \in [0,1) \;,\; t^- \in \RR \; \bigg\}
 \odot (\ZZ, \cdot) \quad. \label{Norm3b}
\end{align}
\end{subequations}
(\ref{Norm3a}) says that the isometry group of the cylinder $C^2_r$
contains translations only; (\ref{Norm3b}) expresses the fact that the
semigroup extension of the isometry group of the cylinder is given by
the discrete Lorentz transformations (\ref{coord5}). A faithful matrix
representation of $eI(C^2_r)$ is obtained from (\ref{CoordAct1}) using
lightlike coordinates (\ref{coord1}),
\begin{equation}
\label{faithsemi2}
 (t,\Lambda_m) = \left( \begin{array}{cc} \Lambda_m & t \\ 0 & 1
 \end{array} \right) = \left( \begin{array}{cc|c} m & 0 & t^+ \\ 0 &
 \frac{1}{m} & t^- \\[5pt] \hline \rule{0pt}{10pt} 0 & 0 & 1
 \end{array} \right) \quad.
\end{equation}
(\ref{faithsemi2}) acts on lightlike coordinates according to
\begin{equation}
\label{faithsemi3}
 (t,\Lambda_m) \left( \begin{mat}{1} & x^+ \\ &x^- \end{mat} \right) =
 \left( \begin{mat}{1} & m\, x^+ + t^+ \\[5pt] & \frac{1}{m}\, x^- + t^-
 \end{mat} \right) \quad.
\end{equation}
We see that these transformations are volume-preserving, since
translations and Lorentz transformations are so, and the
transformations $\Lambda_m$ are still {\it locally injective}; in
particular, they are injective on the tangent spaces. Furthermore, it
is obvious that $\Lambda_m$ is injective on each of the strips
$\left[0, \frac{r}{m} \right) \times \RR$, $\left[ \frac{r}{m},
\frac{2r}{m} \right) \times \RR$, $\ldots$ , separately, but since
each strip is mapped by $\Lambda_m$ onto the whole interval $[0,1)$,
the inverse image $\Lambda_m^{-1}(x^+, x^-)$ of any point $(x^+, x^-)
\in C^2_r$ contains $m$ elements. Yet each of the $\Lambda_m$ is
perfectly metric-preserving,
\begin{equation}
\label{metpres1}
 \Lambda_m^T\, h\, \Lambda_m = h \quad,
\end{equation}
and injective on tangent spaces, as mentioned above. These
transformations therefore should qualify as isometries in an
"extended" sense. What is more, even though these transformations are
not globally injective, they still preserve the cardinality of the
cylinder $C^2_r$, since each of the strips $\left[ \frac{r}{m},
\frac{2r}{m} \right)$, etc., has the same cardinality as the total
interval $[0,1)$.

On the covering space $\RR^2_1$, transformations $\Lambda_m$ are
indeed one-to-one, and any two reference frames {\it on the covering}
$\RR^2_1$ which are related by such a transformation must be regarded
as physically equivalent. This leads to an important question: Should
reference frames on the compactification $C^2_r$ be regarded as
equivalent if they are related by a semigroup transformation
$\Lambda_m$ (\ref{coord5})? Let us examine the consequences of such an
assumption for a scalar field:

\section{Classical scalar fields living on $C^2_r$ \label{Scalar}}

Let us first clarify what we mean by a real scalar field on the
compactification, by comparing it with the definition of an
$O(1,1)$-scalar $\phi$ on the covering space $\RR^2_1$: In $\RR^2_1$
we have a set of equivalent frames which are mutually related by
Lorentz transformations $\Lambda$. In each of these frames, an
observer defines a single-valued field with respect to his coordinate
chart by assigning  $\RR^2_1 \ni X \mapsto \phi(X) \in \RR$ a single
number to each point $X$. Consider any two equivalent frames with
coordinates $X$ and $X' = \Lambda X$, and call the respective field
values $\phi(X)$ and $\phi'(X')$. If these field assignments are
related by
\begin{equation}
\label{scalar1}
 \phi'(X') = \phi(\Lambda^{-1} X'\,) \quad,
\end{equation}
for all $X' \in \RR^2_1$, and then for all pairs of equivalent frames,
we refer to the collection of assignments $\phi, \phi', \ldots$ as an
$O(1,1)$ scalar field.

If we now try to carry over this scenario to the compactified space
$C^2_r$ we see that we run into certain troubles: The set $O(1,1)$ is
reduced to a discrete set (\ref{coord5}) which moreover is now only a
semigroup, so that we cannot operate definition (\ref{scalar1}) which
involves inverses of $\Lambda$. However, (\ref{scalar1}) may be
rewritten in the form
\begin{equation}
\label{scalar2}
 \phi'(\Lambda X) = \phi(X) \quad,
\end{equation}
for all $X \in C^2_r$, and $\Lambda = \Lambda_m$ now. Using
(\ref{faithsemi3}) this can be written explicitly as
\begin{equation}
\label{scalar3}
 \phi'\left(m\, x^+, \frac{1}{m}\, x^- \right) = \phi(x^+, x^-) \quad.
\end{equation}
Thus, if $\phi(X)$ is a single-valued field in frame/chart $X$, {\bf
and} if we adopt the assumption that frames on $C^2_r$ related by
transformations (\ref{coord5}) are physically equivalent, then
frame/chart $X'$ must see a {\bf single-valued} field $\phi'(X')$
obeying (\ref{scalar3}). So, if $x^+$ ranges through
$[0,\smallf{r}{m})$, then $x^{\prime +}$ covers $[0,r)$ once; if $x^+$ ranges
through $[\smallf{r}{m}, \smallf{2r}{m} )$, then $x^{\prime +}$ already has
covered $[0,r)$ twice, and so on. Thus, if $\phi'$ is supposed to be
single-valued, then $\phi$ must be {\bf periodic} with period $r/m$ on
the interval $[0,r)$ in the first place. But then this argument must
hold for arbitrary values of $m$. Barring pathological cases and
focusing on reasonably smooth fields $\phi$ we conclude that $\phi$
must be constant on the whole interval $[0,r)$! As a consequence of
(\ref{scalar3}), each of the equivalent observers reaches the same
conclusion for his field $\phi'$. The statement that the scalar field
$\phi$ is independent of the lightlike coordinate $x^+$ is therefore
"covariant" with respect to the set of $\Lambda_m$-related observers
on $C^2_r$.

It is interesting to see that this fact makes the field $\phi$
automatically a solution to a massless field equation if $\phi$ was
assumed to be massless in the first place: To see this, rewrite the
d'Alembert operator $\Box = -h^{ab} \d_a \d_b$ in lightlike
coordinates using (\ref{coord2}),
\begin{equation}
\label{scalar4}
 \Box = 2\, \d_+ \d_- = 2\, \d_- \d_+ \quad.
\end{equation}
The equation of motion for a massless scalar field is $\Box \phi = 0$,
which, on account of (\ref{scalar4}), is seen to be satisfied
automatically by all scalars which obey the consistency condition
$\d_+ \phi = 0$ as discussed above. On the other hand, a massive
scalar is inconsistent with the peculiar geometry on $C^2_r$ as
manifested in the semigroup structure of the isometry group, since the
massive Klein-Gordon equation $(\Box + \mu_b^2) \phi = 0$ leads
necessarily to $\mu_b^2\, \phi = 0$ for all scalars $\phi$ compatible
with the semigroup structure on $C^2_r$. Let us repeat this important
point:

{\it Any single-valued field $\phi$ on $C^2_r$ which transforms as a
scalar with respect to the extended isometry group $eI(C^2_r)$ is
necessarily a solution to a massless Klein-Gordon equation}.

We can arrive at the same conclusion, and learn more, if we study
plane-wave solutions to the massive Klein-Gordon equation:
\begin{equation}
\label{plw1}
 \phi(x^0, x^1) \sim \cos\big(kx^1 - \omega_k x^0 + \delta \big)
 \quad, \quad \omega_k = \sqrt{ k^2 + \mu_b^2} \quad.
\end{equation}
Here we have started by assuming the general case, so that the
dispersion law on the right-hand side of (\ref{plw1}) is again that of
a massive field, but we shall arrive at the conclusion for
masslessness as above presently. Expressed in lightlike coordinates
eq. (\ref{plw1}) becomes
\begin{equation}
\label{plw2}
 \phi(x^+, x^-) \sim \cos \bigg\{ \frac{1}{\sqrt{2}} \left( k-
 \omega_k \right) x^+ - \frac{1}{\sqrt{2}} \left( k+ \omega_k \right)
 x^- + \delta \bigg\} \quad.
\end{equation}
This quantity must be independent of $x^+$, which is equivalent to
saying that
\begin{equation}
\label{plw3}
 \mu_b = 0 \quad, \quad k = |k| > 0 \quad.
\end{equation}
In other words, only the {\it right-propagating} modes are admissible,
while left-propagating ones, $k<0$, are inconsistent with the
$eI(C^2_r)$-geometry on the compactification. The admissible modes, up
to constant phase shifts, and expressed in $x^0 x^1$-coordinates, then
look like
\begin{subequations}
\label{cf5}
\begin{equation}
\label{cf5a}
\begin{aligned}
 \cos k(x^1-x^0) & = \Re f_k(x) \quad, \quad f_k(x) = e^{ik x^1 - i k
 x^0} = f_k(x^1-x^0) \quad,
\end{aligned}
\end{equation}
so that any field $\phi$ may be decomposed into
\begin{align}
 \phi(x^1-x^0) & = \int\limits_0^{\infty} \frac{dk}{4\pi k}
 \sqrt{\hbar_C}\; \bigg\{ a_k\, f_k(x) + a_k\ad\, f_k^*(x) \bigg\}
 \quad, \quad x = (x^0, x^1) \quad, \label{cf5b}
\end{align}
\end{subequations}
where we have introduced the $SO(1,1)$-invariant integration measure
$dk/4\pi k$. At this point, $\hbar_C$ is just a real numerical factor
with dimension of an action which, in the quantum theory, may be
identified with the Planck quantum of action on the cylinder $C^2_r$.

From (\ref{cf5a}) we immediately see that parity transformation or
time-reversal do not map admissible solutions into admissible
solutions; therefore, both of these symmetries are broken. However, a
combined parity-time-reversal is admissible, as it maps a mode
(\ref{cf5a}) into itself. This is clearly consistent with the findings
in (\ref{coord4A}, \ref{coord5}), where the combined transformation
$\C{P} \C{T}$ was seen to be an isometry of the metric, but $\C{T}$ or
$\C{P}$ separately were not.

Now, what about the dependence of $\phi$ on the coordinate $x^-$? We
can rewrite (\ref{scalar3}),
\begin{equation}
\label{scalar5}
 \phi'(x^{\prime -}) = \phi(m\, x^{\prime -}) \quad, \quad x^{\prime
 -} \in \RR \quad.
\end{equation}
Thus, an observer using chart $X' = \Lambda_m\, X$ sees a "contracted"
version of the field, since, on an interval $[0,\smallf{L}{m})$,
$x^{\prime -}$ has already covered the same information with respect to
$\phi'$ as $x^-$ in the interval $[0,L)$ with respect to $\phi$. More
precisely, the Fourier transforms $\phi(k)$ and $\phi'(k')$ of the
fields $\wt{\phi}(X)$ and $\wt{\phi}'(X')$ are related by
\begin{equation}
\label{scalar6}
 \wt{\phi}'(k') = \frac{1}{m}\, \wt{\phi}\left( \frac{k'}{m} \right)
 \quad,
\end{equation}
hence the frequency spectrum of the scalar field is "blue-contracted",
thus "more energetical", for the $X'$-observer. Suppose that the field
$\phi$ is such that it tends to zero at infinity. Then in the
"infinite-momentum" limit $m \rightarrow \infty$ we see a field with
spatial dependence
\begin{equation}
\label{scalar7}
 \phi'(x^{\prime -}) \; = \; \left\{ \begin{array}{ccl} 0 & , &
 x^{\prime -} \neq 0 \quad, \\[8pt] \phi(0) & , & x^{\prime -} = 0
 \quad. \end{array} \qquad \right\}
\end{equation}
This result is noteworthy: We have started out with a field $\phi$
which was supposed to transform under the semigroup transformations as
a scalar. This transformation behaviour has led us to conclude that
the field must be right-propagating, and must not depend on the
coordinate $x^+$. We have derived this independence by assuming only
that 1.) the lightlike direction $x^+$ is compactified, with no
condition on the length $r$ of the compactified interval; and 2.) that
the semigroup transformations (\ref{coord5}) continued to make sense
as maps between physically equivalent reference frames, which allowed
us to define fields with scalar transformation behaviour under the
semigroup (\ref{coord5}). On the other hand, we must keep in mind that
the transformations $\Lambda_m$ no longer comprise a group, but only a
semigroup with the property (\ref{coord6}) that the succession of two
such transformations can ever go only into {\bf one} direction, namely
towards $m \rightarrow \infty$. In our view this points out the
infinite-momentum frame $m \rightarrow \infty$ as something preferred!
The preference is manifested in the fact that the set of Lorentz
transformations (\ref{coord5}) now has an {\it inherent order}, given
by
\begin{equation}
\label{order1}
 \Lambda_m \prec \Lambda_{m+1} \prec \cdots \quad,
\end{equation}
which follows the order $m < m+1 < \cdots$ of the indices $m$. This
order reflects the fact that a transformation $Y = \Lambda_m X$
between two different frames cannot be undone (after all, it is not
globally injective). This is in stark contrast to the usual case, for
example, the relation between two different frames on the covering
space $\RR^2_1$, where both frames can be reached from each other by
an appropriate Lorentz transformation. We see that the ordering
(\ref{order1}) clearly points out the "infinite-momentum" frame at $m
\rightarrow \infty$ as a {\it preferred frame}. This preferred frame
is an immediate consequence of admitting the semigroup transformations
(\ref{coord5}) to be part of our physical reasoning on the spaces
$C^2_r$. Some consequences of these preferred frames for Kaluza-Klein
theories will be presented in section \ref{KaluzaKlein} below. First,
however, we must study the transformation behaviour of vectors and
covectors under the semigroup transformations (\ref{coord5}).

\section{Vectors and covectors on $C^2_r$ \label{Vectors}}

We can extend the reasoning leading to the scalar transformation law
(\ref{scalar3}) to vector and covector fields on $C^2_r$. Consider a
vector field $V = V^+ \d_+ + V^- \d_-$ on $C^2_r$; then $V$ must
transform under $\Lambda_m$-transformations as $V'(\Lambda_m X) =
\Lambda_m V(X)$, or
\begin{equation}
\label{vector3}
 \left[ \begin{array}{c} V^{\prime +}(m x^+, \frac{1}{m} x^-) \\[5pt]
 V^{\prime -}(m x^+, \frac{1}{m} x^-) \end{array} \right] = \left[
 \begin{array}{c} m\, V^+(x^+, x^-) \\[5pt] \frac{1}{m}\, V^-(x^+,
 x^-) \end{array} \right] \quad.
\end{equation}
As in the case of scalar fields, single-valuedness of the component
fields requires that the fields cannot depend on $x^+$.

Now we consider $1$-forms $\omega = \omega_+\, dx^+ + \omega_-\,
dx^-$, denoting components by row vectors $(\omega_+,
\omega_-)$. Their transformation behaviour is obtained from
\begin{equation}
\label{forms1}
 \omega = \omega_+\, dx^+ + \omega_-\, dx^- = \omega'_+\, dx^{\prime
 +} + \omega'_-\, dx^{\prime -} \quad,
\end{equation}
such that
\begin{equation}
\label{forms2}
 \bigg(\, \omega'_+(\Lambda_m X)\,, \; \omega'_-(\Lambda_m X) \,
 \bigg) = \bigg(\, \omega_+(X)\,,\; \omega_-(X)\, \bigg) \left(
 \begin{mat}{2} & \frac{1}{m} && 0 \\ & 0 && m \end{mat} \right) =
 \left( \omega_+, \omega_- \right)\, \Lambda^{-1}_m \quad.
\end{equation}
Again, we conclude that the fields cannot depend on $x^+$. This latter
feature is clearly general, and applies to tensor fields of arbitrary
type $m\choose n$, since it arises from the transformation of the
arguments $X$ of the tensor fields, rather than the transformation of
the tensor components.

\section{Consequences for six-dimensional Kaluza-Klein theories}
\label{KaluzaKlein}

So far we have studied the consequences of admitting semigroup
transformations to connect different physical frames on the
cylindrical spacetimes $C^2_r$. Now we want to examine these
transformations within a greater framework: We envisage the case of a
product spacetime $M \times C^2_r$, where the external factor $M$ is a
standard 3+1-dimensional Minkowski spacetime with metric $\eta$, while
the internal factor is comprised by our cylindrical space $C^2_r$ with
metric $h_{ab} = \diag(-1,1)$. We use coordinates $(x^{\mu}, x^a) =
x^A$ on the product manifold such that $x^{\mu}, \mu = 0, \ldots, 3$,
and $x^a, a=4,5$, are coordinates on the external Minkowski space and
the internal cylinder, respectively. In the absence of gauge fields,
the signature of the 6-dimensional metric is $(-1,1,1,1,-1,1)$. When
off-diagonal metric components are present, we denote the metric as
\begin{equation}
\label{met1}
 \hat{g} = \left( \begin{array}{c|c} g_{\mu\nu} & A_{\mu a} \\[5pt]
 \hline A_{a \nu} & h_{ab} \end{array} \right) \quad,
\end{equation}
or explicitly,
\begin{equation}
\label{met2}
 \hat{g} = g_{\mu\nu}\, dx^{\mu} \otimes dx^{\nu} + A_{a\mu}\, \big(
 dx^a \otimes dx^{\mu} + dx^{\mu} \otimes dx^a \big) + h_{ab}\, dx^a
 \otimes dx^b \quad.
\end{equation}
Here, $a,b,\ldots$ run over indices $+$ and $-$ on the internal
dimensions; while Greek indices range over the external dimensions $0,
\ldots 3$ on $M$. The submetric $g_{\mu\nu}(x^{\rho}, x^a)$ of the
external spacetime may turn out to depend on the internal coordinates
$x^a$, so all we may demand is that on the 3+1-dimensional submanifold
we are aware of, i.e. for $x^a = 0$, we have $g_{\mu\nu}(x^{\rho}, 0)
= \eta_{\mu\nu}$.

Let us now apply a semigroup transformation (\ref{coord5}) to
(\ref{met2}),
\begin{equation}
\label{met3}
 \hat{g} = g_{\mu\nu}\, dx^{\mu} \otimes dx^{\nu} + A_{a\mu} \left(
 \Lambda_m^{-1} \right)^a_{\ms b} \big( dy^b \otimes dx^{\mu} +
 dx^{\mu} \otimes dy^b \big) + h_{ab}\, dy^a \otimes dy^b \quad,
\end{equation}
where $y^a = \left( \Lambda_m\right)^a_{\ms b}\, x^b$, and the
internal part of the metric remains invariant, since $\Lambda_m$
preserves the metric $h_{ab}$. Comparison with (\ref{forms2}) shows
that the off-diagonal elements $A_{a\mu}$ transform like covectors on
the internal manifold; we therefore conclude that they must be
independent of the $x^+$-coordinate on $C^2$. Hence,
\begin{equation}
\label{met4}
\begin{aligned}
 A'_{+\mu}(y^-) & = \frac{1}{m}\, A_{+\mu}\left(\, m y^- \, \right)
 \quad, \\
 A'_{-\mu}(y^-) & = m\, A_{-\mu}\left(\,m y^- \, \right) \quad,
\end{aligned}
\end{equation}
where, for sake of simplicity, we have temporarily suppressed the
dependence of the gauge potentials $A_{a\mu}$ on the spacetime
coordinates $x^{\mu}$. We now assume that the off-diagonal elements
satisfy
\begin{equation}
\label{met5}
 \lim\limits_{x^- \rightarrow \pm \infty} A_{a\mu}(x^-) = 0
 \quad.
\end{equation}
Then the same arguments given in (\ref{scalar5}, \ref{scalar7}) lead
to the conclusion that, in the "infinite-momentum" limit $m
\rightarrow \infty$, the $A_{+\mu}$-fields vanish, while the
$A_{-\mu}$-fields appear localized around the point $x^- =0$, but
acquire an infinite amplitude. This latter feature is less
catastrophic than it might appear, though: There is no inherent length
scale on the internal manifold which must be preserved, and so nothing
can stop us from applying a conformal rescaling
\begin{equation}
\label{met6}
 \left( \begin{mat}{1} &y^+ \\ &y^- \end{mat} \right) \rightarrow
 \left( \begin{mat}{1} &z^+ \\ &z^- \end{mat} \right) = \left(
 \begin{mat}{1} &y^+ \\ e^{\beta}\; &y^- \end{mat} \right) \quad,
\end{equation}
which is to accompany the limit $m \rightarrow \infty$ in such a way
as to compensate for the factor of $m$ in front of
$A_{-\mu}$. Explicitly, the metric (\ref{met3}) in the coordinates
(\ref{met6}) becomes
\begin{equation}
\label{met7}
\begin{aligned}
 \hat{g} & = \eta_{\mu\nu}\, dx^{\mu} \otimes dx^{\nu} + \\[8pt]
 & + A_{+\mu} \; \frac{1}{m}\; \big( dz^+ \otimes dx^{\mu} + dx^{\mu}
 \otimes dz^+ \big) + \\[8pt]
 & + e^{-\beta}\; A_{-\mu}\; m\; \big( dz^- \otimes dx^{\mu} +
 dx^{\mu} \otimes dz^- \big) + \\[8pt]
 & + e^{-\beta}\, h_{ab}\, dz^a \otimes dz^b \quad.
\end{aligned}
\end{equation}
Hence, if we perform the limits $\beta \rightarrow \infty$ and $m
\rightarrow \infty$ simultaneously by setting $e^{\beta} = m$, then
\begin{equation}
\label{met8}
 \hat{g}\; \xrightarrow{\quad m\rightarrow \infty \quad}\;
 \eta_{\mu\nu}\, dx^{\mu} \otimes dx^{\nu} + A_{-\mu}(x^{\nu}, z^-)\,
 \big( dz^- \otimes dx^{\mu} + dx^{\mu} \otimes dz^- \big) + 0 \cdot
 dz^- \otimes dz^- \quad.
\end{equation}
Eq. (\ref{met8}) is a very interesting result: As seen from the
preferred, infinite-momentum, frame, the field content of the original
metric (\ref{met1}) is that of a Kaluza-Klein theory defined on a flat
$(3+1)$-dimensional external Minkowski space times a one-dimensional
factor (the $z^-$-direction) along which the metric is zero, while the
$z^+$-direction has completely vanished out of sight. In a more
technical language, we obtain an effective $5$-dimensional {\it
non-compactified} \cite{OverduinWesson1997} Kaluza-Klein theory. The
associated metric for $z^- = 0$ is
\begin{equation}
\label{limit1}
 \hat{g} = \left( \begin{array}{c|c} \eta_{\mu\nu} & A_{\mu -} \\[5pt]
 \hline A_{- \nu} & 0 \end{array} \right) \quad.
\end{equation}
In contrast to previous Kaluza-Klein theories, the metric
(\ref{limit1}) is {\it degenerate} in the absence of fields;
furthermore, it may be expected that the signature of the overall
metric $\hat{g}$, now including gauge potentials, will depend on the
particular field configuration represented by the $A_{-\mu}$: The
characteristic equation of the metric (\ref{met1}) is
\begin{equation}
\label{charac1}
 \left( \lambda -1\right)^2\, \Big\{ \lambda^3 - \lambda\, \big(
 \v{A}^2 + A_0^2 + 1 \big) - \v{A}^2 + A_0^2 \Big\} = 0 \quad,
\end{equation}
where we have abbreviated $A_{\mu} \equiv A_{-\mu}$. Two eigenvalues
are always equal to one. If the vector potential satisfies $A_{\mu}
A^{\mu} = 0$ then a zero eigenvalue exists, and the metric is
degenerate even in the presence of fields $A_{\mu}$. We have checked
numerically that both cases $\v{A}^2= 0$ and $A_0^2= 0$ can produce
negative, vanishing and positive eigenvalues.

The emergence of a degenerate metric clearly goes beyond the framework
of standard Kaluza-Klein theories, where the non-degeneracy of the
overall metric is usually taken for granted. However, we are inclined
to take the result (\ref{met8}) serious if it can be shown to produce
meaningful physics. To this end we have to contemplate how equations
of motion for the fields $A_{\mu} \equiv A_{-\mu}$ can be
obtained. The point here is that the usual technique of obtaining
dynamical equations for the fields from the Einstein equations for the
total metric $\hat{g}$ no longer works: On account of the fact that
the inverse $\hat{g}^{-1}$ does not exist in the field-free case, a
Levi-Civita connection on the total manifold is not well-defined,
since the associated connection coefficients (Christoffel symbols)
involve the inverse of the metric. It might be possible to produce
equations of motions for the fields $A_{\mu}$ by going back to the
original field multiplet $A_{a\mu}$, before taking the limit of the
infinite-momentum frame on the internal space; computing the Ricci
tensor, and then trying to obtain a meaningful limit $m \rightarrow
\infty$ for the theory even though the metric becomes degenerate in
this limit. This possibility will be explored elsewhere.

\section{Summary \label{Summary}}

The concept of the extended normalizer of a group of isometries leads
to the possibility of semigroup extensions of isometry groups of
compactified spaces. For flat covering spaces which are compactified
over lattices, semigroup extensions become possible when the lattice
contains lightlike vectors. The simplest example is provided by the
family of two-dimensional cylindrical spacetimes with Lorentzian
signature compactified along a lightlike direction; the members of
this family are all isometric to each other. The semigroup elements
acting on these cylinders can be given a natural ordering, which in
turn suggests the existence of preferred coordinate frames, the latter
consisting of infinite-momentum frames related to the canonical chart
by extreme Lorentz transformations. Fields as viewed from the
preferred frame acquire extreme values; in particular, some of the
off-diagonal components of the total metric, regarded as gauge
potential for a field theory on the external Minkowski factor, may
vanish, leaving a field multiplet which is reduced in size and which
no longer depends on one of the coordinates on the internal space. The
infinite amplitude exhibited by the surviving fields can be removed by
conformal rescaling of one of the lightlike directions. The effective
theory so obtained is a Kaluza-Klein theory which is reduced by one
dimension, and has a smaller field content, but which is defined on a
space with degenerate metric. The latter feature is responsible for
the fact that dynamical equations for the fields $A_{\mu}$ cannot be
obtained from the Ricci tensor of the overall metric.

\acknowledgements{Hanno Hammer acknowledges support from EPSRC
grant~GR/86300/01.}


\end{document}